\documentclass{rspublic}

\usepackage{graphicx}
\usepackage{amsmath}
\usepackage{subfigure,wrapfig}

\begin{document}

\title{Doubly-periodic array of bubbles in a  Hele-Shaw cell}
\author{Ant\^onio M.~P.~Silva and Giovani L.~Vasconcelos}
\affiliation{Laborat\'orio de F\'{\i}sica Te\'orica e Computacional, 
Departamento de F\'{\i}sica, Universidade Federal de Pernambuco,
50670-901, Recife, Brazil.}

\label{firstpage}

\maketitle

\begin{abstract}{free-boundary problems; Hele-Shaw flows; bubble dynamics}
Exact solutions are presented for a doubly-periodic array of steadily moving
bubbles in a Hele-Shaw cell when surface tension is neglected.  It
is assumed that the bubbles either are symmetrical with respect to the
channel centreline or have fore-and-aft symmetry, or both, so that the relevant flow
domain can be reduced to a simply connected region. By
using conformal mapping techniques, a general solution with any number
of bubbles per unit cell is obtained in integral form.  Several
examples are given, including solutions for multi-file arrays of bubbles
in the channel geometry and doubly-periodic solutions in an unbounded cell.
\end{abstract}

\maketitle

\section{Introduction}

Since the seminal papers by Saffman \& Taylor (1958) and Taylor \& Saffman (1959) the
motion of bubbles in a Hele-Shaw cell, where the fluids are confined between two closely spaced parallel plates,  has attracted a great deal of
attention. Because the problem is amenable to analytic treatment,
particularly in the case when surface tension effects are neglected,
many exact solutions have been found for both steady and
time-dependent flows. A comprehensive list of references, up to 1998, on Hele-Shaw flows was compiled by S. D. Howison (people.maths.ox.ac.uk/howison/Hele-Shaw).   
Hele-Shaw flows have also recently been recognized as having deep connections with other areas of mathematical physics (Mineev-Weinstein {\it et al.} 2000, Gustafsson \& Vasil'ev 2006). Exact solutions for Hele-Shaw flows are thus of interest not only for their eventual physical significance but also on their own right for they are solutions to a difficult free-boundary problem. 
Some known exact solutions for the motion of bubbles and fingers in a Hele-Shaw cell were briefly reviewed by  Vasconcelos (2001).  
More recently, other exact solutions have been found, such as time-dependent solutions for Hele-Shaw flows
around a wedge  (Cummings 1999, Richardson 2001b,c, Vasconcelos 2007) and
solutions for multiple steady bubbles in a Hele-Shaw cell with no assumed symmetry obtained by Crowdy (2009a,b).

Exact solutions for a periodic stream of bubbles in a Hele-Shaw cell have also been found.
The first of such solutions was obtained by Burgess \& Tanveer (1991) for a stream of bubbles in the channel geometry  with one bubble per unit cell. In the limit that the period is taken  to infinity their solution recovers the  solution originally given by Taylor \& Saffman (1959) for a single bubble in a channel. Exact solutions for a stream of bubbles in an unbounded  Hele-Shaw cell were subsequently found by Vasconcelos (1993). The Burgess-Tanveer solution  was later generalized by Vasconcelos (1994) for a wider class of solutions with an arbitrary number of symmetrical bubbles per unit cell. Here again taking the period to infinity yields solutions with an arbitrary but finite number of symmetrical bubbles in a Hele-Shaw channel (Vasconcelos 2001).   It is worth mentioning that a periodic solution in a Hele-Shaw channel can be extended to the entire plane  by successive reflections at the channel walls thus yielding a doubly-periodic solution in an unbounded cell. 
In this context, exact solutions for doubly-periodic flows in a Hele-Shaw cell are of interest not only from a mathematical viewpoint, for they correspond to a more general situation, but also from a physical perspective because in certain cases they can be realized as a stream of bubbles in a Hele-Shaw channel and hence may (in principle) be susceptible to experimental investigation. A brief experimental study of a stream of bubbles rising in an inclined Hele-Shaw cell was reported by Maxworthy (1986).

In the present paper, we report a rather general class of exact solutions for a doubly-periodic array of steadily moving bubbles in a Hele-Shaw cell.   Our generic solutions are valid in an unbounded Hele-Shaw cell but there is a large subclass of solutions that correspond to a stream of bubbles in a Hele-Shaw channel. To render the problem analytically tractable, we neglect surface tension effects and suppose that the bubbles either are symmetrical with respect to the channel centreline or have fore-and-aft symmetry (or both), so that the flow domain can be reduced to a simply-connected unit cell. By using conformal mapping techniques, a general
solution for any number of bubbles per unit cell is obtained
 in integral form.    Our solution
 represents the most general periodic solution for a stream of bubbles in a Hele-Shaw cell, in the sense that all previously  known periodic solutions are particular cases of the solutions reported here. Furthermore, our solutions include several novel configurations that were not obtained before, such as the case of a staggered two-file array of bubbles in a channel that somewhat resembles the ``zipper"-like flows
 of red cells in capillaries (Sugihara-Seki \& Fu 2005). Also presented are solutions with ``mixed symmetry" in the channel geometry where within a given bubble configuration some bubbles have fore-and-aft symmetry and others are symmetrical about the channel centreline.  Examples are also given of solutions that cannot be restricted to the channel geometry and hence must necessarily be considered in an unbounded Hele-Shaw cell.

%As already indicated above, most of the known exact solutions for Hele-Shaw flows  are  for the case in which surface tension is neglected, an exception being the class of exact solutions with surface tension for a very particular geometry discussed by Kadanoff (1991), Vasconcelos and Kadanoff (1993) and Vasconcelos (1994).  Although neglecting surface tension represents a rather idealised situation, exact solutions in such approximation are nevertheless of great interest not only from a mathematical viewpoint, for they are analytical solutions to a difficult moving boundary problem, but also because they may describe (at least approximately) actual shapes observed in experiments (Saffman \& Taylor 1958, Vasconcelos 2000).  
%Exact solution for  steadily moving  bubbles in a Hele-Shaw cell usually exhibit a characteristic `degeneracy', meaning that fixing the geometrical parameters of the solution does not fix the bubble velocity. This degeneracy is usually  removed by the inclusion of a small amount of surface tension, in which only certain discrete  values of the bubble velocity are allowed (Tanveer 1987, Combescto \& Dombre 1988). Unfortunately, the search for solutions with surface tension has to be performed numerically which renders the problem rather difficult.

\section{Mathematical formulation}

\begin{figure}
\centerline{\includegraphics[width=0.6\textwidth]{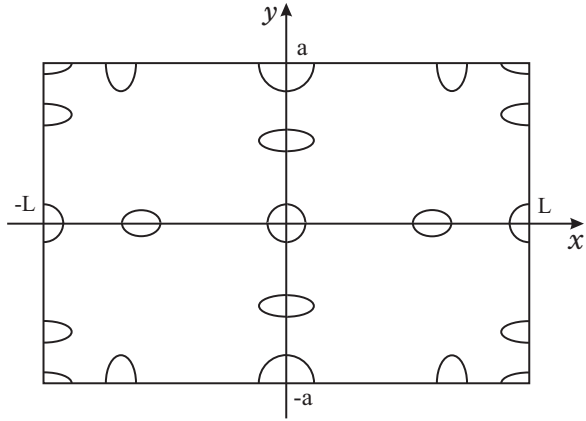}}
\caption{The unit cell for a doubly-periodic array of bubbles in a Hele-Shaw cell.}
\label{fig:1}
\end{figure}

\begin{figure}
\begin{center}
\subfigure[\label{fig2a}]{\includegraphics[width=0.4\textwidth]{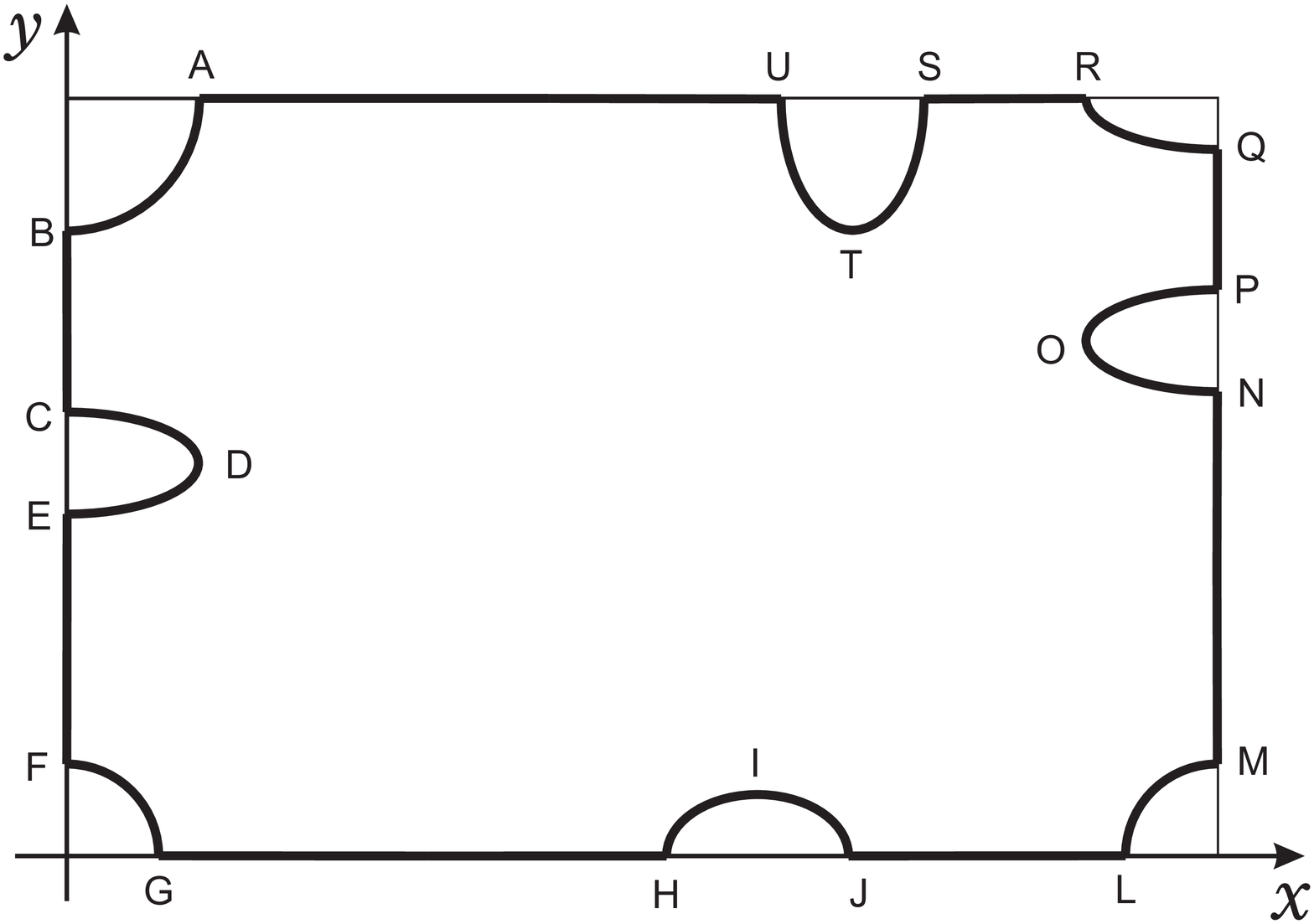}}\qquad 
\subfigure[\label{fig2b}]{\includegraphics[width=0.4\textwidth]{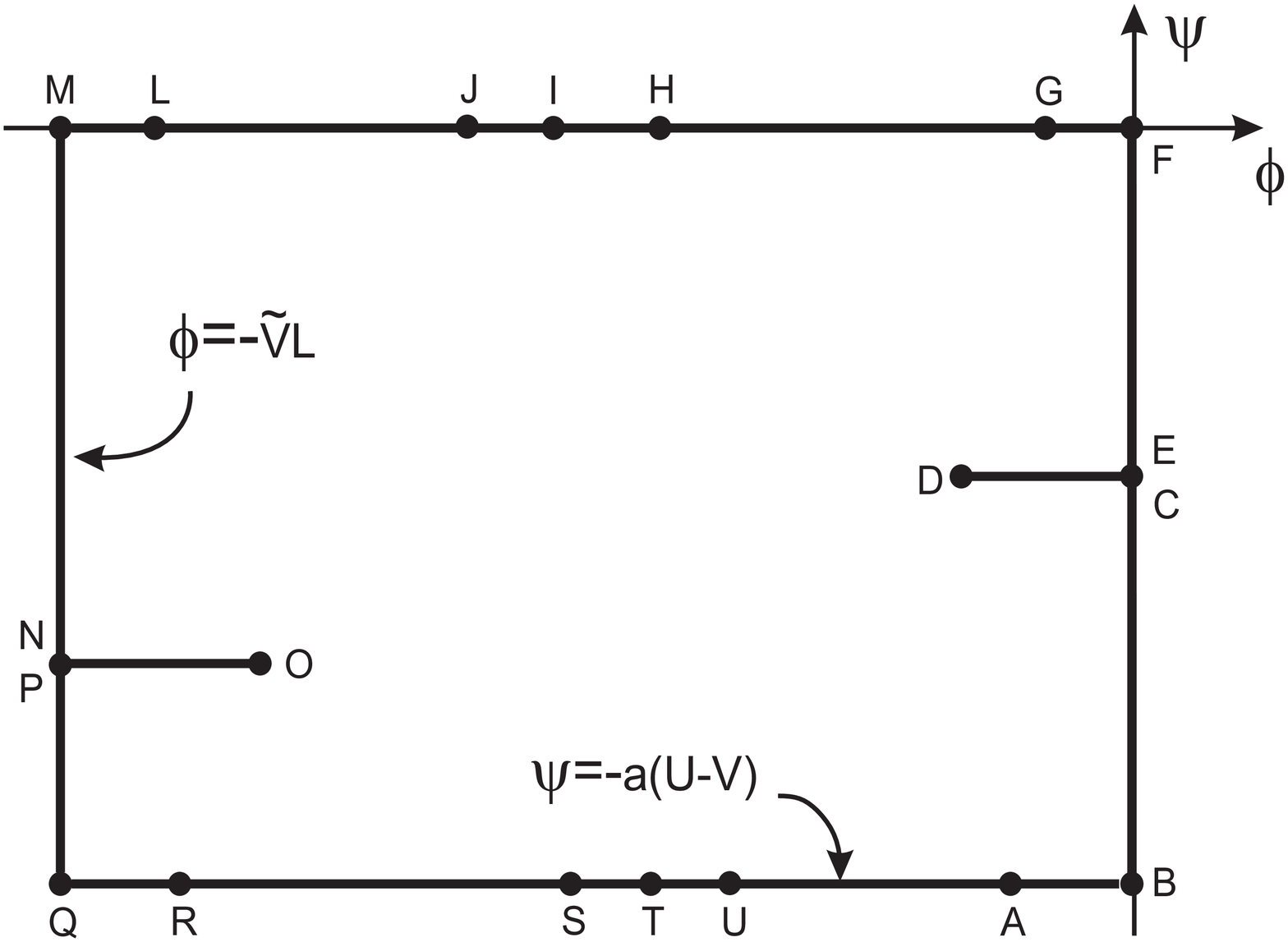}}
\subfigure[\label{fig2c}]{\includegraphics[width=0.4\textwidth]{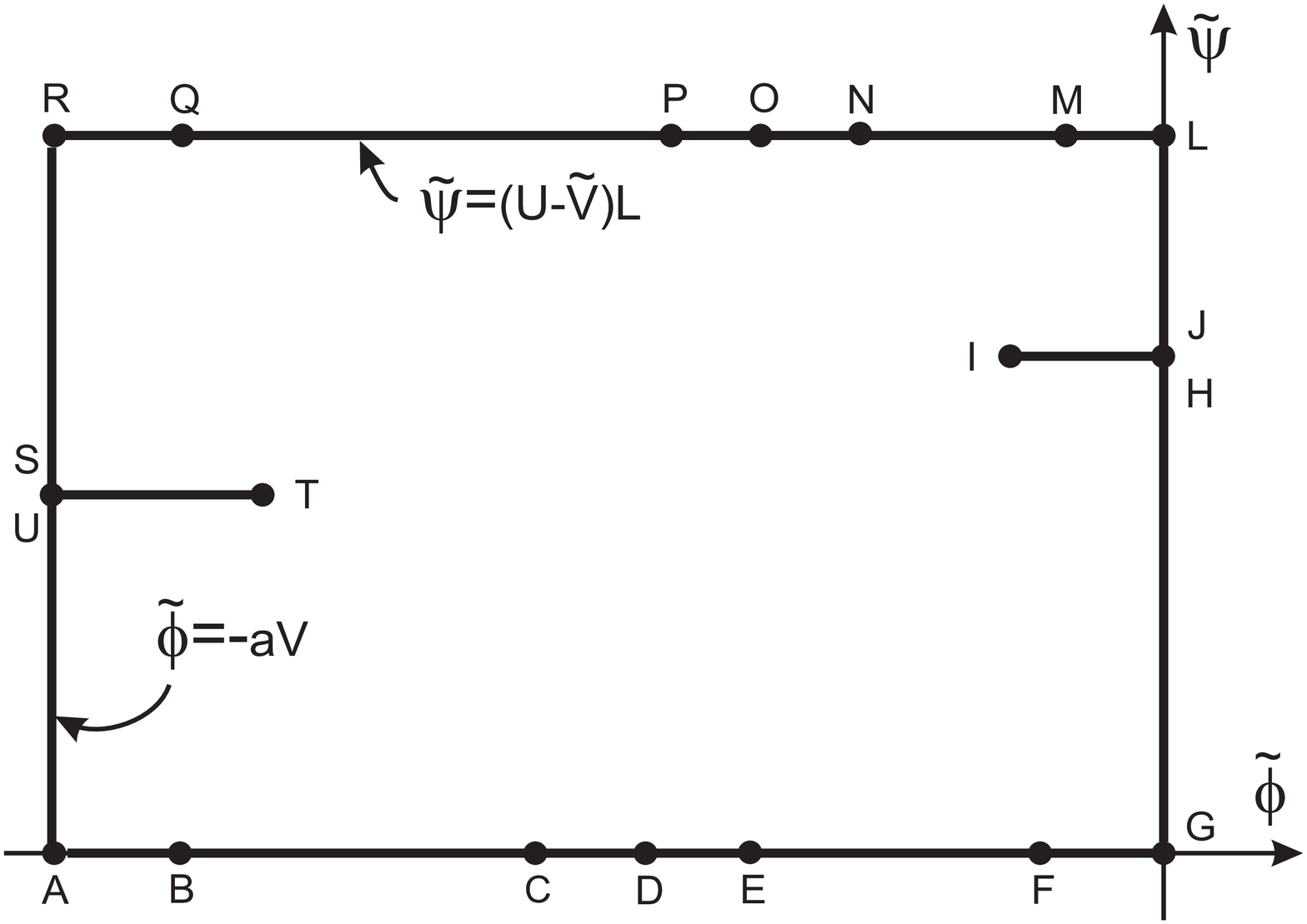}}\qquad 
\subfigure[\label{fig2d}]{\includegraphics[width=0.4\textwidth]{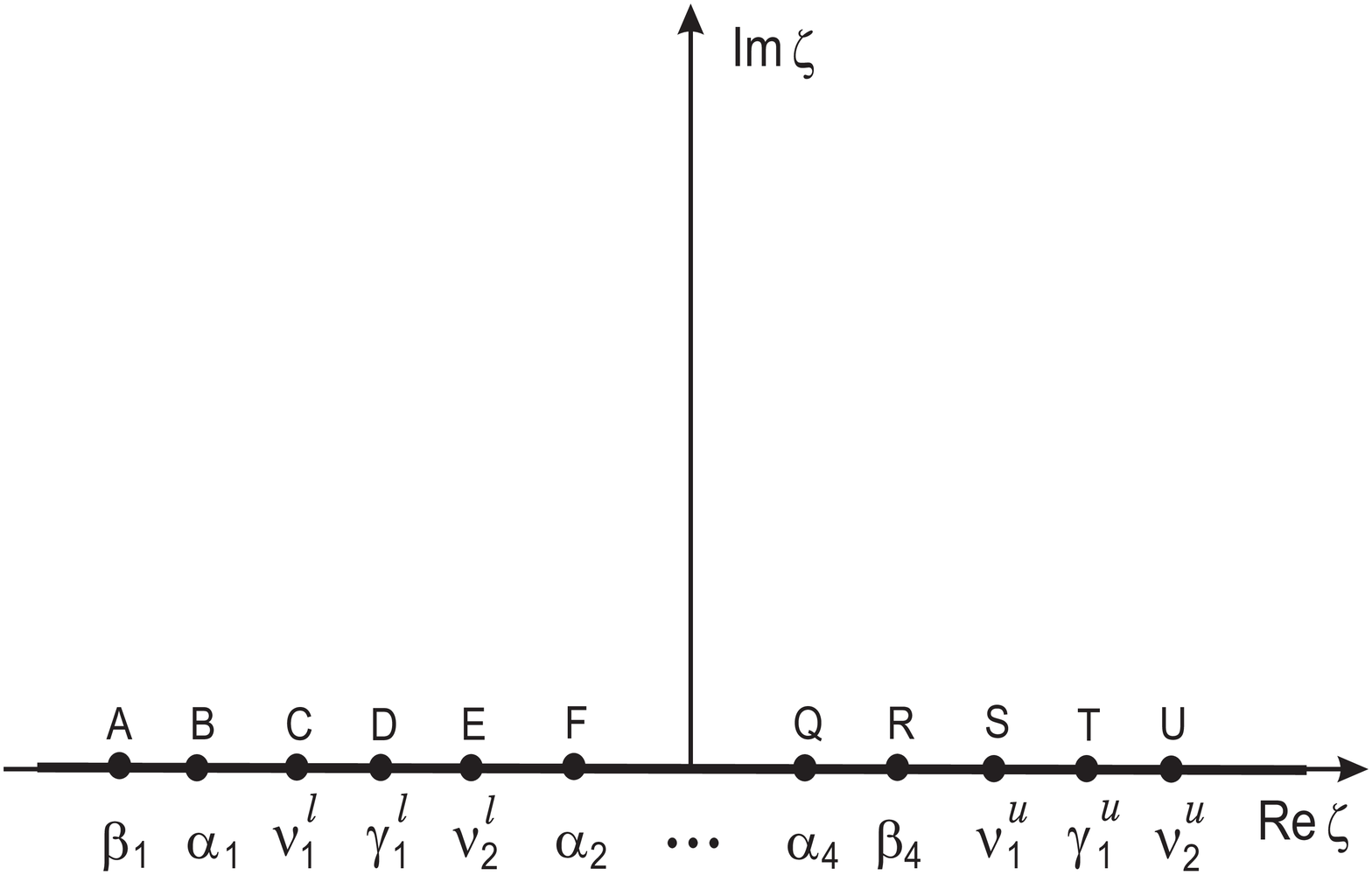}}
\end{center}
 \caption{The reduced unit cell for a doubly-periodic array of bubbles:
 (a) the $z$-plane, (b) the $W$-plane, (c) the $\tilde{W}$-plane, and
 (d) the $\zeta$-plane.}
\label{fig:2}
\end{figure}

We consider the problem of a doubly-periodic array of bubbles moving
with a constant velocity $U$ along the $x$-direction in an unbounded
Hele-Shaw cell. For convenience, we shall work in a reference frame
moving with the bubbles in which the flow is steady and the bubbles
stationary. We assume that the solutions have streamwise
and spanwise periods of $2L$ and $2a$, respectively, and so we reduce
the problem to a rectangular unit cell, as shown in Fig.~\ref{fig:1}. We place the origin of
our system of coordinates at the centre of the unit cell, so that its
upper and lower edges are at $y=\pm a$ and the left and right
edges at $x=\pm L$. We shall assume furthermore that the bubble configuration
is symmetrical with respect to reflections about the following four
axes: $x=0$, $x=L$, $y=0$, and $y=a$. With this assumption we only
need to consider the problem in a reduced unit cell corresponding to
one quarter, say, the upper-right quarter, of the original unit cell; see
Fig.~\ref{fig2a}.

Now let $\cal D$ denote the domain occupied by the fluid within the
reduced unit cell. For simplicity of notation, we shall denote by $\cal
C$ a generic bubble-fluid interface.   As is usual for Hele-Shaw flows, we introduce the complex potential $W(z)
= \phi +\ri \psi$, where $z=x+\ri y$, $\phi(x,y)$ is the velocity potential,   
and $\psi(x,y)$ is the associated stream function. We recall that the
velocity potential (in the moving frame) is given by
$\phi(x,y)=-({b^2 p}/{12\mu})-Ux$, where $b$ is the gap between the
cell plates, $\mu$ is the viscosity, and $p$ is the fluid pressure.
The complex potential $W(z)$ must be
analytic in the fluid domain ${\cal D}$ and satisfy the appropriate
boundary conditions, as we now discuss.

The pressure inside a bubble is assumed to be constant and
surface tension effects are neglected. The complex potential $W$ must then satisfy
the following condition on a fluid-bubble interface ${\cal C}$
\begin{equation}
%W = -Ux +\phi_{j}^i+\ri \psi_{j}^i  \ \ \ {\rm on} \ \ {\cal C}_{j}^i, \label{eq:3e}
W = -Ux +C \ \ \ {\rm on} \ \ {\cal C},
\label{eq:3e}
\end{equation} 
where $C$ is a complex-valued
constant.  (The value of the constant $C$ depends on the  specific bubble one considers and is related to the position of the bubble centroid.) The complex potential must also satisfy specific conditions on the boundaries of the unit cell, as follows.
Given that the
upper and lower edges of our unit cell are streamlines of the flow and  that the left and right edges are equipotentials,
we must
then have
\begin{equation}
\Imag W=  0 \ \  \ {\rm on} \ \ y= 0, \label{eq:3a}
\end{equation}
\begin{equation}
\Imag W=  (V-U)a \ \ \ {\rm on} \ \ y= a, \label{eq:3b}
\end{equation}
\begin{equation}
\Real W= 0 \ \ \ {\rm on} \ \ x= 0, \label{eq:3c}
\end{equation}\begin{equation}
\Real W = -\tilde{V}L   \ \ \ {\rm on} \ \ x= L, \label{eq:3d}
\end{equation}
where Im and Re denote real and imaginary parts, respectively. Here $V$ is the average velocity across the unit cell in the $x$-direction in the lab frame and $\tilde{V}$ is a positive constant whose physical meaning will become clear shortly.  An explicit expression for $\tilde{V}$ in terms
of  $L$, $V$ and the total 
area  $J$ occupied by the bubbles (within a unit cell)  can be obtained. We refer
the reader to Burgess and Tanveer (1991) for the details and simply quote the results here:
\begin{equation}
\tilde{V}=U-V+\frac{UJ}{aL}. \label{eq:tilV}
\end{equation}

The complex potential $W(z)$ defined above can be seen as a conformal
mapping from the fluid region in the $z$-plane onto the corresponding
flow domain in the $W$-plane.  In view of
Eqs.~(\ref{eq:3e})--(\ref{eq:3d}), one readily sees that the flow
domain in the $W$-plane corresponds to a rectangle, $-\tilde{V}L < \phi <
0$, $-(U-V)a<\psi<0$, as shown in
Fig.~\ref{fig2b}.  Bubbles placed along the left and right edges of the unit cell in the $z$-plane are mapped onto horizontal slits in the $W$-plane, whereas  bubbles on the streamlines $y=0$ and $y=a$ are mapped onto segments along the lines $\psi=-a(U-V)$ and $\psi=0$, respectively; see Fig.~\ref{fig2b}. Note that the horizontal slits on Fig.~\ref{fig2b}  correspond to the respective halves of the bubbles on the left and right edges that are within the reduced unit cell and hence such slits  always penetrate into the rectangular domain in the $W$-plane.

As discussed in detail elsewhere (Tian \& Vasconcelos 1993, Vasconcelos 2001), steady Hele-Shaw flows have an interesting rotation invariance in the following sense: if a curve $\cal C$ is a solution for a bubble
moving with constant velocity $U$ along the $x$-direction in an
unbounded Hele-Shaw cell, then the curve $\tilde{\cal C}$ obtained
from a rotation of $\cal C$ about the origin by an angle $\alpha$ is
also a solution with the same velocity $U$. Alternatively, one can view the rotated
solution as one in which the bubble itself is kept fixed while its
velocity is rotated by an angle $\alpha$.  
In view of the rectangular geometry of our unit cell, it is  particularly relevant for us here to consider
a rotation by $90^\circ$, where the bubbles move with the same velocity
$U$ as in the original solution but now in the $y$-direction.  In this case, the region occupied by the fluid in the rotated solution is
exactly the same as that for the original problem, the main difference
being that the upper and lower edges of the unit cell which are streamlines in the original solution become equipotentials   in the rotated problem, whereas the left and right edges which were originally equipotentials become streamlines.  

As shown by Vasconcelos (2001),  the
complex potential   $\tilde{W}(z)$  for the
rotated problem (in the corresponding moving frame) is given by
\begin{equation}
\tilde{W}(z)=\ri\left[W(z)+ U z\right] . \label{eq:tilW}
\end{equation}
From (\ref{eq:3e})--(\ref{eq:3d}) and (\ref{eq:tilW}) it then follows
that the rotated complex potential $\tilde{W}(z)$ satisfies the
following boundary conditions:
\begin{equation}
\tilde{W} = -Uy +\ri C\ \ \ {\rm on} \ \ {\cal C}. \label{eq:4e}
\end{equation}
\begin{equation}
\Real\tilde{W}=  0 \ \  \ {\rm on} \ \ y= 0, \label{eq:4a}
\end{equation}
\begin{equation}
\Real \tilde{W}=  -Va \ \ \ {\rm on} \ \ y= a, \label{eq:4b}
\end{equation}
\begin{equation}
\Imag \tilde{W}= 0 \ \ \ {\rm on} \ \ x= 0, \label{eq:4c}
\end{equation}\begin{equation}
\Imag \tilde{W} =(U-\tilde{V})L  \ \ \ {\rm on} \ \ x= L, \label{eq:4d}
\end{equation}
From (\ref{eq:4d}) one now sees that the parameter $\tilde{V}$
corresponds to the average fluid velocity (in the lab frame)
in the $y$-direction of the rotated flow.
The flow domain for the rotated problem in the $\tilde{W}$-plane
corresponds to a rectangle, $-Va < \tilde{\phi} < 0$,
$UL-\phi_0<\tilde{\psi}<0$, with horizontal slits corresponding to
the bubbles placed on $y=0$ and $y=a$, as shown in Fig.~\ref{fig2c}.  [For the same reasons as discussed before, these horizontal slits always penetrate into the bounded domain in $\tilde{W}$-plane.]

Consider now the conformal mapping $z(\zeta)$ from the upper half-$\zeta$-plane
 onto the fluid domain in the $z$-plane, so that the
boundary $\partial {\cal D}$ of the fluid domain in the $z$-plane is mapped onto the
real axis of the $\zeta$-plane, as shown in Fig.~\ref{fig2d}.  In particular,
a bubble ${\cal C}$  will be the image under
$z(\zeta)$ of a corresponding interval ${\cal I}$ on the real-$\zeta$-axis (see below). For definiteness, we also choose to map the point $\zeta=\infty$ to an
arbitrary point on the upper edge between the left-upper corner and the
next bubble; see Fig.~2. 
In the most general configuration, one has one bubble on each corner and an arbitrary number of bubbles on each edge of the cell. To keep track of all the bubbles, let us denote by ${\cal C}^{c}_i$, $i=1,...,4$, the bubble on the $i$th corner, with the corners numbered anti-clockwise starting from the upper left corner. Under the map $z(\zeta)$ a  corner bubble ${\cal C}^{c}_i$ will be image of the
interval ${\cal I}_{i}^c$ on the real-$\zeta$-axis defined by
\begin{equation}
{\cal I}_{i}^c\equiv \left\{\begin{array}{c}(\alpha_i ,\beta_i),  \ \  i=2,4\cr (\beta_i,\alpha_i), \ \, i=1,3.\end{array}\right.
\label{eq:Ic1}
\end{equation}
(The reason for distinguishing between even and odd corners is one of convenience, for with the above choice one has that in the $W$-plane the corners are the images of the points $\zeta=\alpha_i$, whereas in the $\tilde{W}$-plane the corners corresponds to $\zeta=\beta_i$; see Fig.~\ref{fig:2}.) Note that if we want a configuration with no bubble in the $i$-th corner, for a given $i$, we must simply set $\alpha_i=\beta_i$.

Similarly, we shall denote the bubbles on the edges of the unit cell 
by ${\cal C}^{a}_i$,  where the superscript refers to the  corresponding side, with $a=l, r, u, d$, indicating left, right, up and down, respectively, and  $i=1,...,n_a$, where $n_a$ is the number of bubbles on the respective edge. In the $\zeta$-plane a given bubble ${\cal C}^{a}_i$ will correspond to an interval ${\cal I}_{i}^a$ on the real axis defined by
\begin{equation}
{\cal I}_{i}^a\equiv (\nu^a_{2i-1} ,\nu^a_{2i}), \quad i=1,...,n_a.
\label{eq:Ia}
\end{equation}
Owing to the three degrees of freedom allowed by the Riemann's
mapping theorem, we could arbitrarily fix the values of three parameters, chosen among $\alpha_i$, $\beta_i$ or $\nu_i^a$. At
this stage, however, it is best to 
consider all such quantities as free parameters, deferring to the examples the moment when we shall need to fix the appropriate parameters.

With some abuse of notation let us now write
\[
W(\zeta)\equiv W(z(\zeta)) \quad {\rm and} \quad \tilde{W}(\zeta)\equiv
\tilde{W}(z(\zeta)),\] so that the functions 
$W(\zeta)$ and $\tilde{W}(\zeta)$ can be viewed as the conformal
mappings from the upper half-$\zeta$-plane onto the flow domains in
the $W$- and $\tilde{W}$-planes, respectively.
 It then follows from (\ref{eq:tilW}) that the mapping
 $z(\zeta)$ can be written as
\begin{equation}
z(\zeta)= -\frac{1}{U} \left[W(\zeta) +\ri\tilde{W}(\zeta)\right].
\label{eq:fz}
\end{equation}
Once the mappings  $W(\zeta)$ and $\tilde{W}(\zeta)$ are known, the bubble interface ${\cal C}^{a}_i$ can then be readily computed from (\ref{eq:fz}) by setting $\zeta=s$,
for $s \in {\cal I}_{i}^a$.
%\begin{equation}
%x_{i}^{a}(s)+\ri y_{i}^{a}(s)=z(s), \quad s \in {\cal
%I}_{i}^a.
%\label{eq:xs}
%\end{equation}

Summarizing our procedure so far, we have reduced our original
free-boundary problem to the much simpler problem of obtaining two
conformal mappings, namely, $W(\zeta)$ and $\tilde{W}(\zeta)$, from the upper half-$\zeta$-plane onto  respective 
rectangular domains in the $W$- and $\tilde{W}$-planes.  As we will
see shortly, such mappings can be easily obtained from the
Schwarz-Christoffel formula (Carrier {\it et al.} 1983).  It
should also be noted that in previous solutions that use conformal
mapping (see, e.g. Tanveer 1987), once the mapping $W(\zeta)$
from the chosen domain in the $\zeta$-plane to the flow domain in the
$W$-plane is known, then the mapping $z=f(\zeta)$ is constructed
explicitly so as to satisfy the appropriate boundary conditions. Our
method has the advantage of making this step rather straighforward by
identifying the function $\tilde{W}(z)=i[W(z)+Uz]$ with the complex
potential for the rotated problem, so that solving for the mapping
$\tilde{W}(\zeta)$ then completes the solution.  Before
presenting our general solutions, however, it is worthwhile to point out some
special features of the solutions for the
the case ${U}=2V$.

Solutions with $U=2V$ are special in the sense that solutions for any
$U>V$ can be generated by a proper rescaling of the  solutions for the former case
(Vasconcelos 2001).  
To show this, we first note that
from inspection of the flow domain in the $W$-plane [Fig.~\ref{fig2b}] it follows that 
the mapping $W_U(\zeta)$, for given $a$ and any $U>V$, can be obtained from  the
corresponding mapping $W_{2V}(\zeta)$, for $U=2V$, by the relation
\begin{equation}
W_U(\zeta) =  \frac{U-V}{V}\, W_{2V}(\zeta).
\label{eq:PhiU}
\end{equation}
In particular,  the velocity $\tilde{V}$ [see Fig.~\ref{fig2b}] for both cases
 are related  by
\begin{equation}
\tilde{V}_U=\frac{U-V}{V}\, \tilde{V}_{2V}. \label{eq:phi_u0}
\end{equation}
Similarly, inspection of the flow domain in the $\tilde{W}$-plane [see Fig.~\ref{fig2c}]
reveals that
\begin{equation}
\tilde{W}_U (\zeta) = \tilde{W}_{2V}(\zeta),
\label{eq:SigU}
\end{equation}
with
\begin{equation}
(U-\tilde{V}_{U})L_U=(2V-\tilde{V}_{2V})L_{2V}. \label{eq:uv}
\end{equation}
Inserting (\ref{eq:phi_u0}) into (\ref{eq:uv}) yields the  half-period $L_U$ in terms of the  corresponding value $L_{2V}$ for the case $U=2V$:
\begin{equation}
L_U=\frac{V(2V-\tilde{V}_{2V})}{UV-(U-V)\tilde{V}_{2V}}L_{2V}. \label{eq:LU}
\end{equation}

It then follows from (\ref{eq:fz}), (\ref{eq:PhiU}) and
(\ref{eq:SigU}) that for any $U>V$ the mapping $z_U(\zeta)$
%with the same set of parameters $\{\nu_j^i\}$ as the solution for $U=2V$, 
can
be written in terms of the solutions for $U=2V$ as
\begin{equation}
z_U(\zeta)=-\frac{1}{U}\left[\left(\frac{U}{V}-1\right)W_{2V}(\zeta)+\ri\tilde{W}_{2V}(\zeta)\right],
\end{equation}
which can be more conveniently rewritten as 
\begin{equation}
z_U(\zeta)=-\frac{1}{2V}\left[(1+\rho)W_{2V}(\zeta)+\ri(1-\rho)\tilde{W}_{2V}(\zeta)\right],
\end{equation}
where
\begin{equation}
\rho=1-\frac{2V}{U},
\end{equation}
with $\rho\in(-1,1)$. 
This result thus shows that solutions for any ${U}>V$ can be obtained
 by an appropriate rescaling of the solution with
${U}=2V$ (Vasconcelos 2001).  In view of this
property, we shall only consider the case $U=2V$ in the remainder of
the paper. Without loss of generality we will also set $V=1$ and $a=1$ henceforth.

\section{General solutions}

As indicated in Fig.~\ref{fig2b}, the flow domain in the $W$-plane is a rectangle
with $n_l$ ($n_r$) horizontal slits corresponding to the bubbles on the left (right) 
side of the unit cell.  This domain can be viewed as a degenerate
polygon and so the corresponding mapping $W(\zeta)$ can be
obtained from the Schwarz-Christoffel formula (Carrier {\it et al.} 1983). One finds that $W(\zeta)$ is determined by
\begin{equation}
 \frac{dW}{d\zeta}= K\,
 \frac{ \prod\limits_{i=1}^{n_l} (\zeta -\gamma_i^{l})
       \prod\limits_{j=1}^{n_r} (\zeta -\gamma_j^{r})} 
 { 
\left[\prod \limits_{i=1}^4(\zeta-\alpha_i)
\prod \limits_{j=1}^{2n_l}\left(\zeta
 - \nu_j^{l}\right)\prod \limits_{k=1}^{2n_r}\left(\zeta -
 \nu_k^{r}\right)\right]^{1/2}} \, ,
\label{eq:Phi}
\end{equation}
where $K$ and $\gamma_i^a$ are real-valued constants to be determined later. 
%In view of the boundary condition
%(\ref{eq:3b}), we have written explicitly the dependence on $U$ and
%$V$ of the overall factor appearing in (\ref{eq:Phi}), where the
%constant $C$ no longer has any dependence on the velocities; see below
%for the determination of $C$. 
In the $\tilde{W}$-plane the flow domain is of the same form as
that in the $W$-plane, namely, a rectangle with $n_d$ ($n_u$)
horizontal slits emanating from the right (left) side of the
rectangle. By the same reasoning as before, we obtain that the map
$\tilde{W}(\zeta)$ is determined by
\begin{equation}
 \frac{d\tilde{W}}{d\zeta}= -\ri\tilde{K} \,
 \frac{ \prod\limits_{i=1}^{n_d} (\zeta -\gamma_i^{d})
         \prod\limits_{j=1}^{n_u} (\zeta -\gamma_j^{u})
      } 
      { \left[\prod \limits_{i=1}^4(\zeta-\beta_i)
	\prod \limits_{j=1}^{2n_d}\left(\zeta
 	- \nu_j^{d}\right)\prod \limits_{k=1}^{2n_u}\left(\zeta -
 	\nu_k^{u}\right)\right]^{1/2}
      } \, ,
\label{eq:Sigma}
\end{equation}
where again $\tilde{K}$ and $\gamma_i^a$ are real-valued constants to be determined. [Note that in (\ref{eq:Phi}) and (\ref{eq:Sigma}) a given constant $\gamma_i^a$ is the pre-image of the point on the
bubble ${\cal C}_{i}^{a}$ that lies furthest away from the respective edge; see Fig.~\ref{fig:2}.]

For calculation purposes it is convenient to expand the numerator in
(\ref{eq:Phi}) and (\ref{eq:Sigma}), and so we write
\begin{equation}
 \frac{dW}{d\zeta}= K\,
 \frac{  \sum\limits_{j=0}^{n_l+n_r} a_j\zeta^j}
 { 
\left[\prod \limits_{i=1}^4(\zeta-\alpha_i)
\prod \limits_{j=1}^{2n_l}\left(\zeta
 - \nu_j^{l}\right)\prod \limits_{k=1}^{2n_r}\left(\zeta -
 \nu_k^{r}\right)\right]^{1/2}} \, ,
 \label{eq:Phi'}
\end{equation}
\begin{equation}
 \frac{d\tilde{W}}{d\zeta}= -\ri\tilde{K} \,
 \frac{ \sum\limits_{j=0}^{n_d+n_u} b_j\zeta^j }
      { \left[\prod \limits_{i=1}^4(\zeta-\beta_i)
	\prod \limits_{j=1}^{2n_d}\left(\zeta
 	- \nu_j^{d}\right)\prod \limits_{k=1}^{2n_u}\left(\zeta -
 	\nu_k^{u}\right)\right]^{1/2}
      } \, ,
\label{eq:Sigma'}
\end{equation}
where the $a_j$'s and $b_j$'s are real-valued coefficients, with
$a_{n_l+n_r}=b_{n_d+n_u}=1$. [The $a_j$'s and $b_j$'s could of course
be expressed in terms of the $\gamma_i^a$'s but it is more convenient to work directly with the mappings
(\ref{eq:Phi'}) and (\ref{eq:Sigma'}).] 
For a given set of
parameters $\alpha_{i}$, $\beta_{i}$ and $\nu_i^a$, the coefficients $a_j$ and $b_j$ are determined as follows.

First, let us define the quantities $I_{ij}^a$ as
\begin{equation}
I_{ij}^a=\int_{\nu_{2i-1}^a}^{\nu_{2i}^a} \frac{t^j dt}
  {\left[\prod \limits_{k=1}^4|t-\alpha_k|\prod \limits_{m=1}^{2n_l}\left|t
 - \nu_m^{l}\right|\prod \limits_{n=1}^{2n_r}\left|t -
 \nu_n^{r}\right|\right]^{1/2}}  ,
\label{eq:I}
\end{equation}
 for  $a=l, r$, with $i=1,...,n_a$ and $j=0,1,...,n_l+n_r$.  We then note from Fig.~\ref{fig:2} that
the mapping functions $W(\zeta)$ must satisfy the
condition   $W(\nu_{2i-1}^a)=W(\nu_{2i}^a)$, for $a=l, r$,  which in
view of (\ref{eq:Phi'}) and (\ref{eq:I}) yield
\begin{equation}
\sum\limits_{j=0}^{n_l+n_r} I_{ij}^a a_j = 0, \quad {\rm for}  \ \ a=l, r \ \ 
 {\rm and}  \  \ i=1,2,...,n_a. 
\label{eq:21}
\end{equation}
Now recalling that $a_{n_l+n_r}=1$, we see that (\ref{eq:21}) gives a
system of $n_l+n_r$ linear equations for the
$n_l+n_r$ unknown coefficients $a_j$.
Similarly, if we define the quantities
\begin{equation}
J_{ij}^a=\int_{\nu_{2i-1}^a}^{\nu_{2i}^a} \frac{t^jdt }
  {\left[\prod \limits_{k=1}^4\left|t-\beta_k\right|\prod \limits_{m=1}^{2n_d}\left|t
 - \nu_m^{d}\right|\prod \limits_{n=1}^{2n_u}\left|t -
 \nu_n^{u}\right|\right]^{1/2}} \, ,
\label{eq:J}
\end{equation}
for $a=d, u$, with $i=1,...,n_a$ and $j=0,1,...,n_d+n_u$, and use the fact that
$\tilde{W}(\zeta)$ must satisfy the
condition    $\tilde{W}(\nu_{2i-1}^a)=\tilde{W}(\nu_{2i}^a)$,  for $a=d,u$,  we obtain
\begin{equation}
\sum\limits_{j=0}^{n_d+n_u} J_{ij}^a b_j = 0, \quad {\rm for}  \ \ a=u,d \ \ {\rm and}  \  \ i=1,2,...,n_a,
\label{eq:22}
\end{equation}
which represent a system of $n_u+n_d$ linear equations for the
$n_u+n_d$ coefficients $b_j$. Thus, once the set of parameters
$\alpha_{i}$, $\beta_{i}$ and $\nu_i^a$ are given we can readily
obtain the constants $a_j$ and $b_j$ by solving the linear equations
(\ref{eq:21}) and (\ref{eq:22}).

The remaining constants $K$ and
$\tilde{K}$ are determined by the boundary conditions (\ref{eq:3b}) and
(\ref{eq:4b}), respectively. To see this, first note that 
 $K$ can be calculated from the requirement
that as we go from point $B$ to point $F$ in Fig.~\ref{fig2d} the complex potential
$W$  in Fig.~\ref{fig2b} must vary by $\ri$  (recall that we have made $U=2$, $V=1$, and $a=1$), and so we write 
\begin{equation}
\left[W\right]_B^F\equiv W(F)-W(B)=\ri , \label{eq:K}
\end{equation}
which yields
\begin{equation}
K^{-1}=-i\int_{\alpha_{1}}^{\alpha_{2}} \frac{\sum\limits_{j=0}^{n_l+n_r} a_j t^j \, dt}
  {\left[\prod \limits_{i=1}^4(t-\alpha_i)\prod \limits_{j=1}^{2n_l}\left(t
 - \nu_j^{l}\right)\prod \limits_{k=1}^{2n_r}\left(t -
 \nu_k^{r}\right)\right]^{1/2}}  .
\label{eq:C}
\end{equation}
The constant $\tilde{K}$ follows from similar requirement:
\begin{equation}
\left[\tilde{W}\right]_A^G\equiv\tilde{W}(G)-\tilde{W}(A)=1,\label{eq:Ktil}
\end{equation}
which yields 
\begin{equation}
\tilde{K}^{-1}=\int_{\beta_{1}}^{\beta_{2}} \frac{\left|\sum\limits_{j=0}^{n_d+n_u} b_j t^j \,\right| dt}
  {\left[\prod \limits_{i=1}^4|t-\beta_{i}|\prod \limits_{j=1}^{2n_d}\left|t
 - \nu_j^{d}\right|\prod \limits_{k=1}^{2n_u}\left|t -
 \nu_k^{u}\right|\right]^{1/2}} \, .
\label{eq:Cc}
\end{equation}

We have thus seen that the generic solution described in (\ref{eq:Phi'}) and (\ref{eq:Sigma'}) is completely specified by prescribing the set of free parameters $\alpha_{i}$, $\beta_{i}$ and $\nu_{i}^{a}$, from which all other constants appearing in the solution can be calculated. (Recall that on account of the three degrees of freedom allowed by the Riemann mapping theorem we can fix any three of the above free parameters.) We note furthermore that all physical parameters of the solution can be calculated once the mathematical parameters are given. 
For example, the streamwise half-period $L$ can be calculated from the requirement
that as we go from point $F$ to point $M$ in Fig.~\ref{fig2a}, the mapping
$z(\zeta)$ must change by $L$:
\begin{equation}
\left[z\right]_F^M\equiv\left. z\right|_M-\left. z\right|_F=L
%f(\zeta=\nu_0^3)-f(\zeta=\nu_{-1}^2),
\end{equation}
which in view of (\ref{eq:fz}) reads
\begin{equation}
\left[W\right]_F^M+\ri \left[\tilde{W}\right]_F^M=2L,
\end{equation}
so that from the knowledge of $W(\zeta)$ and $\tilde{W}(\zeta)$ one readily obtains $L$. Similarly, the bubble centroids and areas can in principle be obtained by carrying out the appropriate calculations, but we shall not go into these details here.
In the next section we will give specific examples of the generic solutions described above.
%Once the two mapping functions $W(\zeta)$ and $\tilde{W}(\zeta)$
%have been determined,  the interface for bubble ${\cal C}_{i}^{a}$ is
%given in parametric
%form by setting $\zeta=s\in {\cal I}_{i}^{a}$  in (\ref{eq:z}):
%\begin{equation}
%x_i^a(s)+\ri y_{i}^{a}(s)=\frac{1}{2}\left[W(s)+\ri \tilde{W}(s)\right], \quad s\in {\cal I}_{i}^{a}.
%\end{equation}

\section{Examples}

\begin{figure}[t]
\centering
{\includegraphics[width=0.45\textwidth]{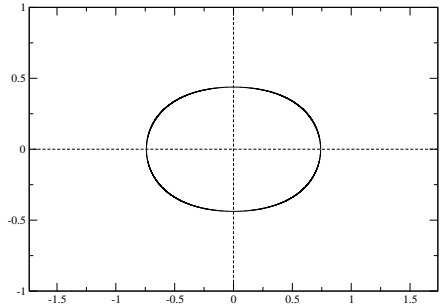}}
\caption{Single-file stream of  bubbles in a Hele-Shaw channel corresponding to the Burgess-Tanveer solution. 
Here the parameters are $\alpha_{2}=-0.6$, $\beta_{2}=0.7$, and $\alpha_{3}=\beta_{3}=0.9$.}
\label{fig:BT}
\end{figure}

\begin{figure}[t]
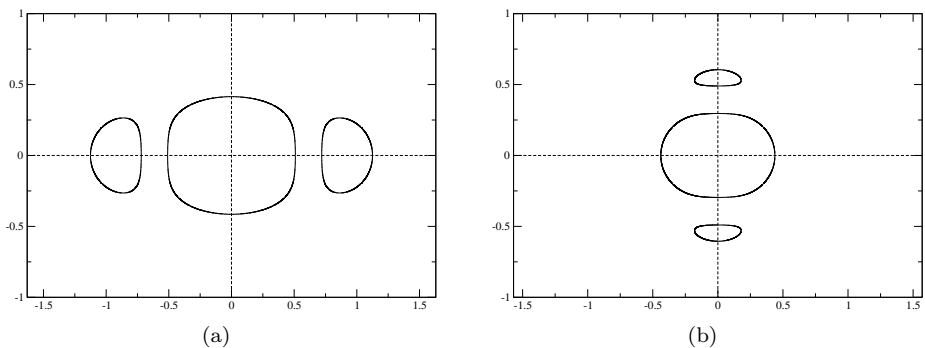

\vspace{0.5cm}
\centering
\subfigure[\label{fig:ch1a}]{\includegraphics[width=0.45\textwidth]{fig4a.eps}}\qquad
\subfigure[\label{fig:ch1b}]{\includegraphics[width=0.45\textwidth]{fig4b.eps}}
\caption{Solutions with (a) centreline symmetry and (b) fore-and-aft symmetry. 
The parameters in (a) are  $\alpha_{2}=-0.9$, $\beta_{2}=-0.5$,
$\nu_{1}^{d}=-0.49$,
$\nu_{2}^{d}=0.4$, and
$\alpha_{3}=\beta_{3}=0.5$; whereas in (b) we have $\nu_{1}^{l}=-0.9$,
$\nu_{2}^{l}=-0.6$, 
$\alpha_{2}=-0.59$, 
$\beta_{2}=-0.2$, and
$\alpha_{3}=\beta_{3}=0.5$.}
\end{figure}

\begin{figure}[t]
\vspace{0.6cm}
\centering
{\includegraphics[width=0.6\textwidth]{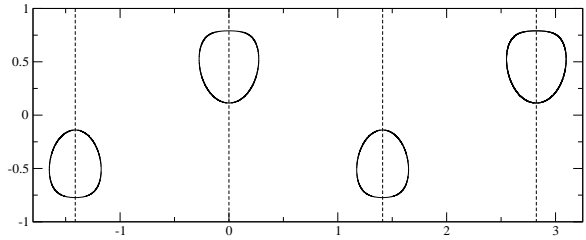}}
\caption{Solution for a two-file array of bubbles in a channel. 
Here the parameters are $\nu_{1}^{l}=-0.9999$, $\nu_{2}^{l}=-0.3$, $\alpha_{2}=\beta_{2}=-0.2$, $\alpha_{3}=\beta_{3}=0.3$, $\nu_{1}^{r}=0.3001$, $\nu_{2}^{r}=0.7$.}
% The vertical lines indicate equipotentials.}
\label{fig:zigzag}
\end{figure}

In this section   we shall consider some particular cases of the generic solution given in the preceding section. We start discussing the situation where we can confine the solutions to a Hele-Shaw channel. After this, we will give an example of solutions that cannot be realized within the channel geometry and hence must be considered in an unbounded Hele-Shaw cell.

\subsection{Periodic bubbles in a channel}

Here we consider the particular and experimentally relevant
case of a periodic array of bubbles moving in a Hele-Shaw channel. 
We suppose the channel walls are at
$y=\pm 1$, so that $y=0$ corresponds to the channel centreline. For this case we set in the generic solution  given in (\ref{eq:Phi'}) and (\ref{eq:Sigma'})
the following conditions 
\begin{equation}
\alpha_{1}=\beta_{1}=-1 , \qquad \alpha_{4}=\beta_{4}=1, \qquad n_{u}=0, \label{eq:nuchann}
\end{equation}
 since there can be no bubble attached to the channel walls. 
 % In this case,  the general solution given in (\ref{eq:Phi'}) and (\ref{eq:Sigma'}) becomes
%\begin{equation}
% \frac{dW}{d\zeta}=
% K \frac{\sum\limits_{j=0}^{n_l+n_r} a_j\zeta^j
%   }
% { \left[(\zeta^2-1)
%\prod \limits_{i=2}^3(\zeta-\alpha_i)
%\prod \limits_{i=1}^{2n_l}\left(\zeta
% - \nu_i^{l}\right)\prod \limits_{i=1}^{2n_r}\left(\zeta -
% \nu_i^{r}\right)
%%\prod \limits_{a=l,r}\prod \limits_{j=1}^{2n_a}\left(\zeta- \nu_j^{a}\right)
% \right]^{1/2}} \, ,
%\label{eq:PhiC}
%\end{equation}
%\begin{equation}
% \frac{d\tilde{W}}{d\zeta}= \ri \tilde{K}\,
% \frac{\sum\limits_{j=0}^{n_d} b_j\zeta^j} 
%      { \left[(\zeta^2-1)\prod \limits_{i=1}^2(\zeta-\beta_i)
%	\prod \limits_{j=1}^{2n_d}\left(\zeta
% 	- \nu_j^{d}\right)\right]^{1/2}} \, ,
%\label{eq:SigmaC}
%\end{equation}
%with the constants $K$, $\tilde{K}$, and $\gamma_{i}^{a}$
%% for $i=1,...,n_{a}$ and$a=l,r,d$, 
%being obtained as described  in the preceeding section.
Thus, for the case of a periodic array of bubbles in the channel geometry the solution 
%described in(\ref{eq:PhiC}) and (\ref{eq:SigmaC})
 is completely specified by
prescribing the following set of free parameters: $\{\alpha_i\}_{i=2,3}$,
$\{\beta_{i}\}_{i=2,3}$, and $\{\nu_{i}^{a}\}_{i=1,...,2n_a}$, with
$a=l,r,d$.  
(Recall that from the third degree of freedom allowed by the Riemann mapping theorem one of these parameters can be kept fixed.) 
We also note that the generic solutions for multiple bubbles in a Hele-Shaw channel presented by Vasconcelos (2001) can be obtained as special case of our periodic solution above in the limit $L\to\infty$. This can be accomplished by simply setting $\beta_{1}=\beta_{2}=-1$ and $\beta_{3}=\beta_{4}=1$ in Fig.~\ref{fig2d}.
In what follows we shall discuss some specific examples of periodic solutions
in a Hele-Shaw channel.

\begin{figure}[t]
\vspace{0.6cm}
\centering
\includegraphics[width=0.6\textwidth]{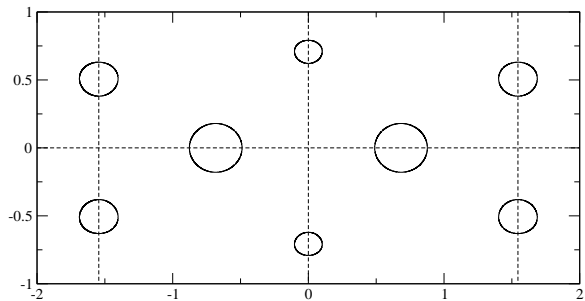}
\caption{Solutions for a multi-file array of bubbles with mixed symmetry.  The parameters are             $\nu_{1}^{l}=-0.97$,
$\nu_{2}^{l}=-0.7$,  $\alpha_{2}=\beta_{2}=-\alpha_{3}=-\beta_{3}=-0.5$, $\nu_{1}^{d}=-0.4$, $\nu_{2}^{d}=0.3$, $\nu_{1}^{r}=0.55$, and $\nu_{2}^{r}=0.9$.}
\label{fig:mixed1}
\end{figure}

\begin{figure}[t]
\vspace{0,5cm}
\centering
\includegraphics[width=0.6\textwidth]{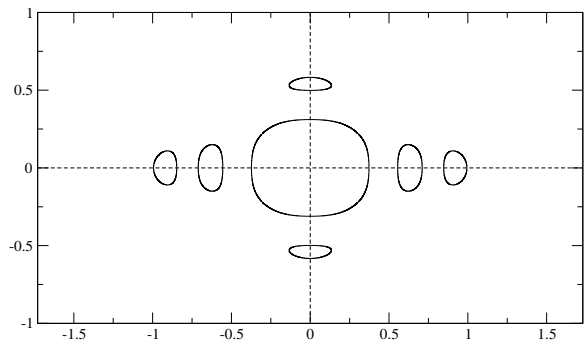}
\caption{Solutions with mixed symmetry including a symmetrical bubble at the centre of the unit cell.  The parameters are $\nu_{1}^{l}=-0.8$, $\nu_{2}^{l}=-0.5$, $\alpha_{2}=-0.48$, $\beta_{2}=-0.1$, $\nu_{1}^{d}=-0.08$, $\nu_{2}^{d}=0.3$, $\nu_{3}^{d}=0.32$,
$\nu_{4}^{d}=0.6$, and $\alpha_{3}=\beta_{3}=0.8$.}   
\label{fig:mixed2}
\end{figure}

We start with the case of a single bubble per unit cell. In the
notation of Fig.~\ref{fig:1} this is achieved by placing a bubble at the lower
left corner of the reduced unit cell and  no other bubble anywhere. In
this case we reproduce the exact solution for an infinite stream
of bubbles in a Hele-Shaw cell originally obtained by  Burgess and
Tanveer (1991) using a rather  different method. For completeness we show in Fig.~\ref{fig:BT} an example of the Burgess-Tanveer solution.
Note, in particular, that in this case the bubble is placed at the centre of the unit cell and therefore has both 
centreline and fore-and-aft symmetry.

As the next example,  we consider the situation in which there are bubbles only either at the lower edge
or at  the left edge of our reduced unit cell. Examples of these two cases are shown in Figs.~\ref{fig:ch1a} and \ref{fig:ch1b}. 
In such cases we recover the class of periodic
solutions with centreline symmetry or fore-and-aft symmetry, respectively, 
obtained earlier by Vasconcelos (1994). It should be noted, however, that in the work by Vasconcelos (1994) no specific example was given, only the general method of solution was presented. In this sense, this is the first time that explicit solutions for such symmetrical configurations have been calculated.

Next we note that novel solutions can be generated if we place bubbles
simultaneously at the left and at the right edges of the reduced
unit cell (and nowhere else).  For example, in Fig.~\ref{fig:zigzag} we show a solution
with two bubbles per unit cell, where each bubble has fore-and-aft
symmetry but they lie in different equipotentials of the flow. Seen in
the laboratory frame, this solution corresponds to a staggered
two-file array of bubbles moving steadily down the channel. It is
interesting to notice that similar zipper-like arrangements are
observed in the flow of red cells in capillaries (Sugihara-Seki \& Fu 2005).

Another type of new solutions includes the case of ``mixed symmetry''
where some bubbles have centreline symmetry and some others have
fore-and-aft symmetry. An example of such case is the multi-file array of bubbles shown in
Fig.~\ref{fig:mixed1},  where there is one file of bubbles symmetrical about the channel centreline, with the other files of bubbles having fore-and-aft symmetry. Another solution with mixed symmetry is shown in Fig.~\ref{fig:mixed2} where in this case there is a symmetrical bubble at the centre of the unit cell.

\subsection{Solutions in an unbounded cell}

\begin{figure}[t]
\vspace{0.5cm}
\centering
{\includegraphics[width=0.5\textwidth]{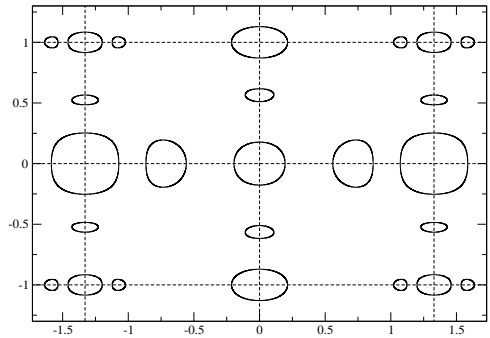}}
\caption{Solution for a doubly-periodic array of bubbles in an unbounded Hele-Shaw cell. Here the parameters are
$\beta_{1}=-1.3$,
$\alpha_{1}=-1$,
$\nu_{1}^{l}=-0.9$,
$\nu_{2}^{l}=-0.6$,
$\alpha_{2}=-0.5$,
$\beta_{2}=-0.4$,
$\nu_{1}^{d}=-0.3$,
$\nu_{2}^{d}=0.3$,
$\beta_{3}=0.32$,
$\alpha_{3}=0.5$,
$\nu_{1}^{r}=0.55$,
$\nu_{2}^{r}=0.8$,
$\alpha_{4}=1$,
$\beta_{4}=1.1$,
$\nu_{1}^{u}=1.12$, and $
\nu_{2}^{u}=1.5$.}
\label{fig:unbounded}
\end{figure}

Here we briefly discuss the case in which the periodic array of bubbles cannot be physically realized within
the channel geometry  and hence must be considered in an unbounded
Hele-Shaw cell. We begin by noting that if there are bubbles placed on the upper and lower
edges of our reduced unit cell, then the solutions cannot be reduced to
the channel geometry as before, and so one must considered an unbounded
Hele-Shaw cell.  An example of such case is given in
Fig.~\ref{fig:unbounded}. 
It is worth mentioning that this type of solution represents the most
general solution for a periodic array of (symmetrical) bubbles in a Hele-Shaw cell, in
the sense that the other types of solution can be obtained by either `decimating' or
`adding' bubbles along any of the edges  of the reduced unit cell. 

\section{Conclusions and Discussion}

We have presented an exact solution for a doubly-periodic array of 
steadily moving bubbles in a Hele-Shaw  cell with an arbitrary number of bubbles per unit cell.  Our solution
 represents the most general periodic solution for bubbles in a Hele-Shaw cell  known to date, in the sense that the previously  known solutions  
for a stream of bubbles in a Hele-Shaw channel are particular cases of the solutions reported here. In addition, our solution includes novel cases of 
periodic arrays of bubbles in a Hele-Shaw channel, such as  multi-file flow of symmetrical bubbles and configurations with mixed symmetry, where some bubbles 
have centreline symmetry while others have fore-and-aft symmetry. 
Examples of a doubly-periodic array of bubbles in an unbounded cell that cannot be restricted to the channel geometry  were also discussed.

In constructing our solutions we assumed that the bubbles either are symmetrical either with respect to the cell centreline or have fore-and-aft symmetry (or both), 
so that the relevant flow domain can be reduced to a simply-connected unit cell. In this context, the existence of periodic solutions  in a Hele-Shaw cell without any 
symmetry requirement, where the unit cell is no longer simply-connected, remains an open and interesting question. Recently, a new methodology to treat Hele-Shaw flows
 in multiply-connected domains has been developed by Richardson (2001a) and Crowdy (2009a,b). In particular, solutions for a finite number of asymmetric steadily-moving
 bubbles have been found both in an infinite Hele-Shaw cell (Crowdy 2009a) and in the channel geometry  (Crowdy 2009b). It is therefore likely that our solutions can be 
generalized to include
the case of a periodic array of asymmetric bubbles. This interesting (and more difficult) problem clearly deserves to be studied further.

Another interesting point to be investigated is the effect of surface tension on the shape and velocity of our periodic array of bubbles. From previous works  on the related 
problem for a single bubble (Tanveer 1986, 1987;    Combescot \& Dombre 1988) as well for the Burgess-Tanveer periodic solution (Burgess \& Tanveer 1991), one expects that the 
`degeneracy'  displayed by our solutions, meaning that fixing the geometrical parameters of the solution does not fix the bubble velocity $U$, will most certainly be removed 
by surface tension. However, it is not clear which particular bubble arrangements, among our large class of solutions, will have counterparts in the case of nonzero surface 
tension. For instance, in the case of a single bubble in a Hele-Shaw channel,  Tanveer  (1987) was unable to find solutions without symmetry  with respect to channel centreline, 
even though such nonsymmetric solutions exist when surface tension is neglected. In light of these results, it appears that periodic solutions with centreline symmetry  are more 
likely to `survive' in the nonzero surface tension scenario. 
%We note however that Tanveer's  analysis (Tanveer 1987) did not rule out the existence of nonsymmetric solution, showing only that  they are difficult to find numerically (i.e., if they exist at all), and so this point remains an open question. In this context, it would be interesting to find out whether the zig-zag solutions shown in Fig.~\ref{fig:zigzag} would have counterparts in the presence of surface tension. 
Unfortunately, however, the search for solutions with surface tension has to be performed numerically---a task that is likely to become increasingly more demanding as the number
 of bubble grows.
%As a final note, we mention that there have been relatively few experimental studies of stream of bubble in a Hele-Shaw cell, 
%an exception reported by Maxworthy (1988). and so it is hoped that the exact solutions reported here might stimulate further experimental investigation of this interesting problem.

\begin{acknowledgements}
This
work was supported in part by the Brazilian agencies FINEP, CNPq, and FACEPE and 
by the special programs PRONEX and CT-PETRO. 
\end{acknowledgements}

\end{document}